\documentclass[pdftex,10pt, letterpaper]{article}
\usepackage{opex3}
\usepackage{graphics}
\usepackage{braket}
\usepackage{amsmath}
\usepackage{amsfonts}
\usepackage{amssymb}
\usepackage{cite}

\usepackage{wasysym}
\usepackage{times}

\renewcommand{\vec}[1]{\mathbf{#1}}

\renewcommand{\Re}{\operatorname{Re}}

\newcommand{\figref}[1]{Fig.~\ref{fig:#1}}
\newcommand{\Figref}[1]{Figure~\ref{fig:#1}}

\newcommand{\Figreftwo}[2]{Figures~\ref{fig:#1} and~\ref{fig:#2}}
\renewcommand{\eqref}[1]{Eq.~(\ref{eq:#1})}

\newcommand{\citeasnoun}[1]{Ref.~\cite{#1}}
\newcommand{\secref}[1]{Sec.~\ref{sec:#1}}

\begin{document}

\title{Bonding, antibonding and tunable optical forces in asymmetric membranes}

\author{Alejandro W. Rodriguez$^{1,2}$, Alexander P. McCauley$^3$,
  Pui-Chuen Hui$^2$, David Woolf$^2$, Eiji Iwase$^2$, Federico
  Capasso$^2$, Marko Loncar$^2$ and Steven G. Johnson$^1$}
\address{$^1$Department of Mathematics, Massachusetts Institute of
  Technology, Cambridge, MA 02139} \address{$^2$School of Engineering
  and Applied Sciences, Harvard University, Cambridge, MA 02138}
\address{$^3$Department of Physics, Massachusetts Institute of
  Technology, Cambridge, MA 02139}

\begin{abstract}  
We demonstrate that tunable attractive (bonding) and repulsive
(anti-bonding) forces can arise in highly asymmetric structures
coupled to external radiation, a consequence of the
bonding/anti-bonding level repulsion of guided-wave resonances that
was first predicted in symmetric systems. Our focus is a geometry
consisting of a photonic-crystal (holey) membrane suspended above an
unpatterned layered substrate, supporting planar waveguide modes that
can couple via the periodic modulation of the holey
membrane. Asymmetric geometries have a clear advantage in ease of
fabrication and experimental characterization compared to symmetric
double-membrane structures. We show that the asymmetry can also lead
to unusual behavior in the force magnitudes of a bonding/antibonding
pair as the membrane separation changes, including nonmonotonic
dependences on the separation. We propose a computational method that
obtains the entire force spectrum via a single time-domain simulation,
by Fourier-transforming the response to a short pulse and thereby
obtaining the frequency-dependent stress tensor. We point out that by
operating with two, instead of a single frequency, these evanescent
forces can be exploited to tune the spring constant of the membrane
without changing its equilibrium separation.
\end{abstract}




\section{Introduction}
\label{sec:intro}

\begin{figure}[t]
\centering
\includegraphics[width=0.5\columnwidth]{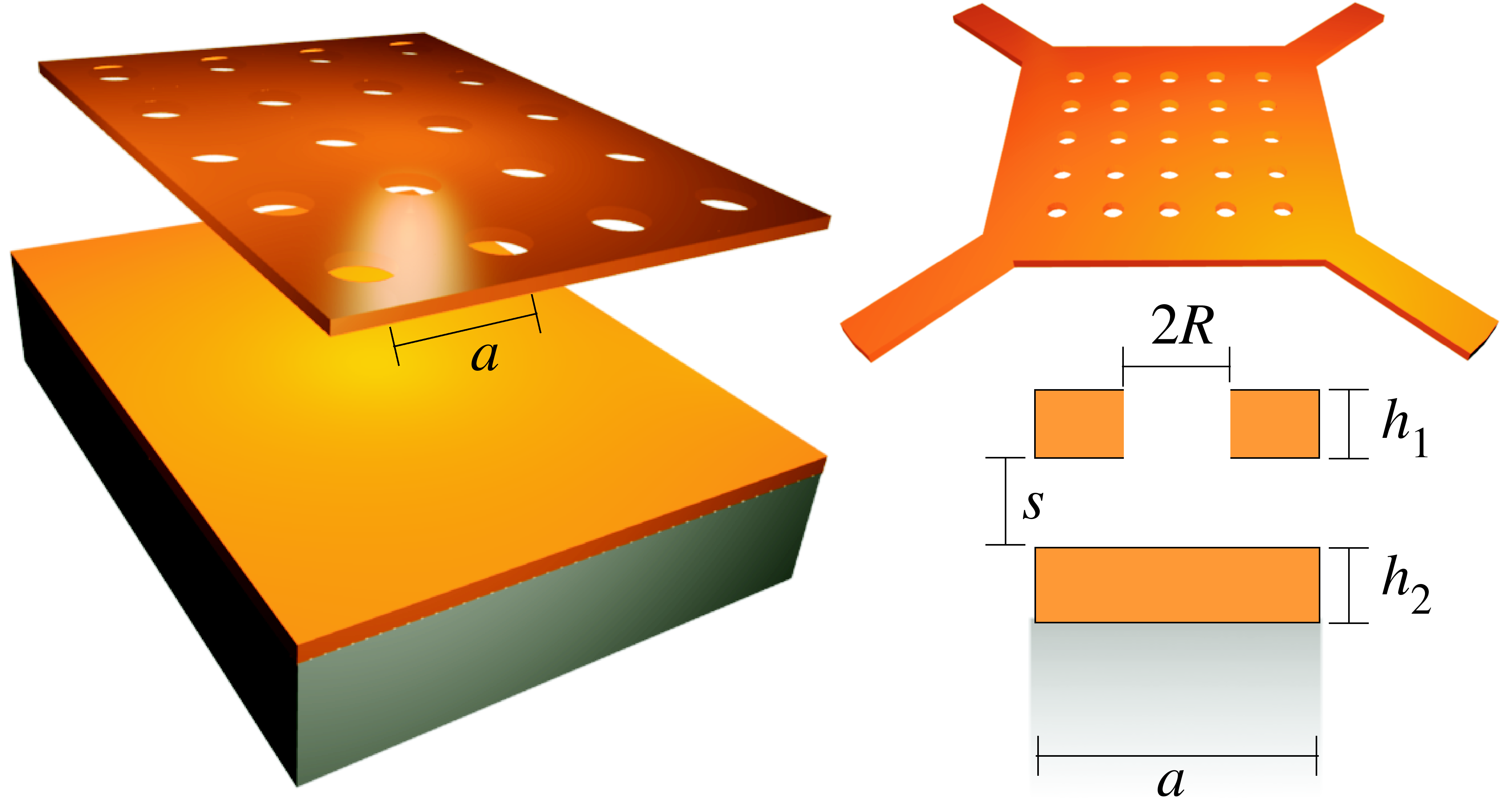}
\caption{Schematic of single-membrane (asymmetric) structure: a
  photonic-crystal (holey) membrane (thickness $h_1 = 0.2a$)
  consisting of a square-lattice of air holes (radius $R =0.2a$) on
  silicon is suspended (separation $s$) on top of an unpatterned
  (homogeneous) silicon slab (thickness $h_2 = 0.2a$) sitting on top
  of a semi-infinite silica substrate. Light is incident on the
  membrane from the normal direction (top).}
\label{fig:fig1}
\end{figure}

The coupling of the electromagnetic field and matter can give rise to
optical forces on otherwise neutral
objects~\cite{Ashkin80,Jackson98,Grier03,Wiederhecker09,Povinelli05},
and optomechanical coupling via radiation pressure or gradient forces
in nanophotonics has been the subject of numerous recent theoretical
and experimental
works~\cite{Kawata96,Vogel03,Metzger04,suh05,Halterman05,Schmidt07,Favero07,JayichSa08,TaniyamaNo08,Anetsberger09,Lu09,Arnold09,LinHu09,Li09,Wiederhecker09,Rosenberg09,YangMo09,Groblacher09,Pernice09,Liu09,Roels09,RohTanabe10,Stomeo10,Aspelmeyer10,LinSch10,Thourhout10,Wiederhecker10}. For
example, two identical (symmetric) waveguides or membranes can attract
or repel depending on the phase and frequency with which they are
excited~\cite{Huang94,
  NgChan05,Povinelli05,TaniyamaNo08,Chan09,WoolfLoncar09}, and this
prediction led to several experimental
realizations~\cite{Wiederhecker09,Rosenberg09,Li09,RohTanabe10}. Here,
we extend this concept to greater generality by demonstrating that
tunable attractive and repulsive forces can also arise in highly
asymmetric structures coupled to external radiation, combining a
suspended holey membrane with an unpatterned layered substrate,
forming two coupled planar waveguides as shown in \figref{fig1}. Such
asymmetric geometries have a clear advantage in ease of fabrication
and experimental characterization compared to symmetric
double-membrane structures (in which two membranes must be both
patterned and suspended).  Although our system is asymmetrical,
repulsive/attractive forces can still be thought of as arising from
the bonding/anti-bonding level repulsion of guided-wave resonances
similar to the phenomenon studied in symmetric
systems~\cite{Povinelli05,Pernice09,Liu09}. In contrast to the
symmetric case, however, it is more difficult to identify precisely
which resonant modes of the large-separation (isolated) waveguides
participate to produce a resonant force, especially at higher
frequencies, since the absence of exact degeneracies means that many
more modes can potentially interact. Unlike symmetric systems,
however, we can obtain bonding/antibonding forces that either increase
or decrease with separation and more generally are nonmonotonic. We
analyze the resonant forces from the perspective of perturbation
theory. In unpatterned multilayer systems, we prove that only
repulsive (not attractive) resonances can occur.  Although
experimental characterization of forces in any membrane system
requires accurate measurement of the membrane displacement, we point
out that the transmission/reflection resonances of the membrane itself
can be used as a sensitive position sensor.  Finally, in contrast to
previous theoretical works that relied primarily on frequency-domain
calculations, we propose a time-domain stress-tensor computational
method that obtains the entire force spectrum from a single
calculation via Fourier transforms of a short pulse.

Optical forces can arise due to radiation pressure or gradient
forces~\cite{Grier03,Kippenberg08}.  Radiation pressure can be thought
of as exchange of momentum between a photon (momentum $\hbar\omega/c$)
and matter, and as a consequence it is easily seen that light with
incident power $P$ exerts a force $F=P/c$ on a planar surface if the
light is 100\% absorbed or a force $F=2P/c$ if it is 100\% reflected.
Therefore, the ratio $Fc/P$ is a useful dimensionless measure of the
strength of an optical force.  Gradient forces, as shown
recently~\cite{Povinelli05}, can arise from the evanescent interaction
between localized optical modes, and the resonant increase in the
field intensity greatly enhances the force for a given input
power~\cite{PovinelliJo05}, so that $|F|c/P \gg 1$.  Such large forces
enable strong, tunable optomechanical
interactions~\cite{Vahala03,Kippenberg:07,Kippenberg08,JayichSa08,Anetsberger09,Groblacher09,Aspelmeyer10},
which have applications such as optical
cooling~\cite{Hansch75,Ashkin80,Metzger04,Favero07,Schliesser08,LinRo09},
optical tweezers and
traps~\cite{Ashkin70,ChuBj86,Phillips88,Grier03,Dholakia05,Corbitt07},
and optical switches~\cite{suh05,Stomeo10}.  Successful demonstrations
of other prominent resonant optomechanical effects include coherent
mechanical oscillation (amplification) of mesoscopic objects with long
vibrational lifetimes, and demonstrations of the photon--phonon
strong-coupling regime via dressed optical
states~\cite{Kippenberg08,Aspelmeyer08,Aspelmeyer10,Akram10}.

We showed in~\citeasnoun{Povinelli05} that if two identical waveguides
or resonant cavities are bought together, the mutual interaction of
these (degenerate) resonances or guided modes can induce a splitting
of the modes into pairs characterized by attractive and repulsive
mechanical forces, analogous to the well-known bonding/anti-bonding
states formed by the level splitting (avoided crossings) of
interacting degenerate states in quantum systems~\cite{Landau:QM}.  In
submicron-scale photonic devices, these forces are strong enough to
yield displacements and other mechanical effects that have been
observed
experimentally~\cite{Wiederhecker09,Rosenberg09,Li09,RohTanabe10}.
Furthermore, the frequency and/or phase of the optical excitation can
be controlled to yield tunable optomechanical effects, even switching
the sign of the force from attractive to repulsive.  There are two
ways to induce repulsive and attractive interactions, depending on
whether the incident power comes in the form of a guided mode or
external radiation (the focus of this paper).  First, one can inject
light parallel to the membranes/waveguides, exciting guided modes that
propagate along the waveguides and interact evanescently.  In this
case, the sign of the interaction is controlled by the relative phase
of the modes in the two
waveguides~\cite{Povinelli05,TaniyamaNo08,Chan09,WoolfLoncar09}.  A
similar effect occurs by controlling the relative phase of two coupled
cavities (e.g. microspheres or microdisks), but in this case the
bonding/anti-bonding resonances also have different frequencies that
can be used to control the sign of the
force~\cite{PovinelliJo05,Rosenberg09,Rakich07}.  It is also possible
to design asymmetric waveguide/cavity structures (e.g. a dielectric
waveguide and a microdisk resonator) with repulsive and attractive
interactions as long as both structures support propagating modes,
again with light incident along the waveguide
direction~\cite{Wiederhecker09,Thourhout10}. (Coupled propagating
modes in asymmetric geometries were also recently shown to lead to
non-monotonic forces~\cite{MaPovinelli10}.) Second, one can shine
light perpendicular to the membranes; if the membranes are perforated
by periodic holes (or any other periodic modulation), normally
incident radiation can couple via diffraction to guided-mode
resonances within the membranes, which again couple evanescently.  In
this case, because the bonding/anti-bonding resonances have different
frequencies, the sign of the force can be controlled by the frequency
of the incident light.  (By considering lateral shifts, one can also
obtain lateral forces and other effects~\cite{Liu09}.)  As the
periodic modulation (e.g. the hole radius) is made smaller, the
lifetime (or quality factor $Q$) of the guided-wave resonances
increases~\cite{Liu09}, the resonant fields become stronger (intensity
$\sim Q$), and thus the resonant forces become stronger $\sim Q$
(albeit narrower in bandwidth $\sim 1/Q$).  This force enhancement is
ultimately limited only by losses (absorption or scattering from
finite size or disorder).  Another limitation is that the narrow
bandwidth translates into a sensitive dependence of the force on the
separation of the two membranes (since the resonant frequency shifts
with separation).

As indicated in \figref{fig1}, this paper considers the case of
normal-incidence light on a suspended membrane, but in contrast to
previous works, we only have a single suspended membrane
(e.g. silicon) over a solid unpatterned substrate.  The substrate must
still support a guided mode of its own in order to obtain forces by
evanescent coupling, and in our case this is achieved by a layered
structure of a higher index layer on a lower-index substrate
(e.g. silicon on silica).  As in previous works, the membrane is
periodically patterned to couple resonantly with normal-incidence
light, and thanks to evanescent coupling this periodicity is also
sufficient to couple light to guided resonances in the unpatterned
substrate.  Because the two waveguides in our case are so different,
however, the interaction is more complicated than the degenerate
level-splitting that occurs in symmetric systems, leading to
nonmonotonic force dependences as well as transitions in the sign of
the force. The separation and frequency dependence of the force in
membrane structures has been previously exploited for achieving a
variety of optomechanical effects.  For example, one can obtain
mechanical oscillators with dynamically tunable ``spring constants,''
even flipping the sign to yield an unstable equilibrium and mechanical
bistability, with a tunable relative strength of the linear and
nonlinear terms (with arbitrarily strong relative nonlinearity
possible if the linear term is
canceled)~\cite{Vahala03,Kippenberg:07,Kippenberg08,JayichSa08,Anetsberger09,Groblacher09,Aspelmeyer10}. Here,
we point out that by operating with light consisting of two
frequencies, rather than a single frequency, one can tailor the spring
constant of the membrane without changing its mechanical equilibrium
separation. Furthermore, we argue that, with appropriate design, it
should be possible for repulsion to dynamically activate at small
separations, creating a feedback mechanism for combating stiction
arising from other forces
(e.g. electrostatic~\cite{FrankPa10,Perhaia10} or
quantum~\cite{Serry98,hochan1} interactions).

\section{Computational Method}
\label{sec:method}

Previous numerical methods for computing optical forces have mainly
focused on frequency-domain approaches, with some
exceptions~\cite{Zhang04,Gauthier05:OE,Benito08}. Given two objects
separated by a distance $d$, the force on one of the objects can be
computed in one of at least two ways: One approach involves computing
the derivative of the eigenfrequencies $\omega$ of two membranes
separated by distance $s$ as a function of $s$, which can be related
to the force via the relation $F = 1/\omega\,
d\omega/ds$~\cite{Povinelli05,PovinelliJo05}, where $F > 0$
corresponds to repulsion. \citeasnoun{LiPernice09} generalized this
formula to handle the case of resonances with finite lifetimes $\gg
2\pi\omega$, rigorously justifying our earlier
suggestion~\cite{PovinelliJo05} that the change in lifetime with
separation has a negligible (higher-order) contribution to the force
on resonance. This approach is problematic, however, in general
circumstances where there may not be well-defined resonant modes with
negligible loss, nor does it include cases where there is a
superposition of the resonant mode with other waves (e.g. light from
an external source).  Another approach involves computing the force
via an integral $\oiint_S \langle \mathbf{T} \rangle \cdot d\vec{S}$
of the time-average Maxwell stress tensor
\begin{equation}
\langle T_{ij} \rangle = \frac{1}{2} \Re \left[ \varepsilon_0 \left(
  E_i E^*_j - \frac{1}{2} \sum_k E_k E^*_k \right) \right. 
  + \left. \mu_0 \left( H_i H^*_j - \frac{1}{2} \sum_k H_k H^*_k \right) \right]
\end{equation}
around some bounding surface $S$ lying in vacuum~\cite{Jackson98}.
For resonant modes with negligible loss, $\langle T_{ij} \rangle$ can
be computed directly from an eigenmode calculation~\cite{Pernice09}.
More generally, including the case where the fields are excited from
an external source whose effect must be included from some
time-harmonic current source $\vec{J}(\vec{x})e^{-i\omega t}$, one can
solve a set of linear equations for the resulting time-harmonic fields
$\vec{E}$ and $\vec{H}$ by a variety of methods (e.g. finite elements
or differences in the frequency domain, or transfer-matrix
methods)~\cite[appendix D]{JoannopoulosJo08-book}, and use these
fields to compute
$T_{ij}$~\cite{Halterman05,Yannopapas08,XiaoChan08,RohTanabe10,AntonoyiannakisPe10}.
This approach, however, has the drawback that if the force at many
frequencies $\omega$ is desired, one must perform many separate
calculations (one for each $\omega$).

If a broad-band force spectrum is desired, an attractive alternative
is to compute the stress tensor via the Fourier transform of a short
pulse in the time domain (e.g. finite-difference time-domain,
FDTD~\cite{Taflove00}), yielding the entire frequency spectrum at
once. Here, one simply evolves Maxwell's equations in response to a
pulse source [e.g. $\sim \vec{J}(\vec{x}) \delta(t)$] in time,
accumulating the discrete-time Fourier-transform [$\tilde{f}(\omega)
  \sim \sum_n f(n\Delta t) e^{i\omega t} \Delta t$] of both the
electric $\vec{E}$ and magnetic $\vec{H}$ fields over the
stress-tensor surface $S$, and at all desired frequencies
$\omega$~\cite{Farjadpour06,Oskooi10:Meep}.  These Fourier-transformed
fields then yield the stress tensor and hence the force.  Of course,
the force must be normalized in some way, and here the dimensionless
$Fc/P$ normalization is very convenient.  One simply does a separate
calculation, with no structure (vacuum), to compute the
Fourier-transformed incident fields and hence the incident power
$P(\omega)$.  Dividing $F(\omega)c/P(\omega)$ yields the dimensionless
force spectrum, where all arbitrary normalization factors (e.g. the
incident pulse spectrum or the normalization of the Fourier transform)
have canceled.  (Matters are more complicated in a nonlinear system,
of course.)

In what follows, we exploit our FDTD approach to compute forces on the
geometry of \figref{fig1}. All of the subsequent calculations were
performed using MEEP, a free FDTD simulation software package develped
at MIT~\cite{Oskooi10:Meep}. We find that discretization errors coming
from our finite resolution of 40~pixels/$a$ affect the computed force
spectra by no more than a few percent.

\section{Membrane Forces}
\label{sec:forces}

\begin{figure}[t!]
\includegraphics[width=1.0\columnwidth]{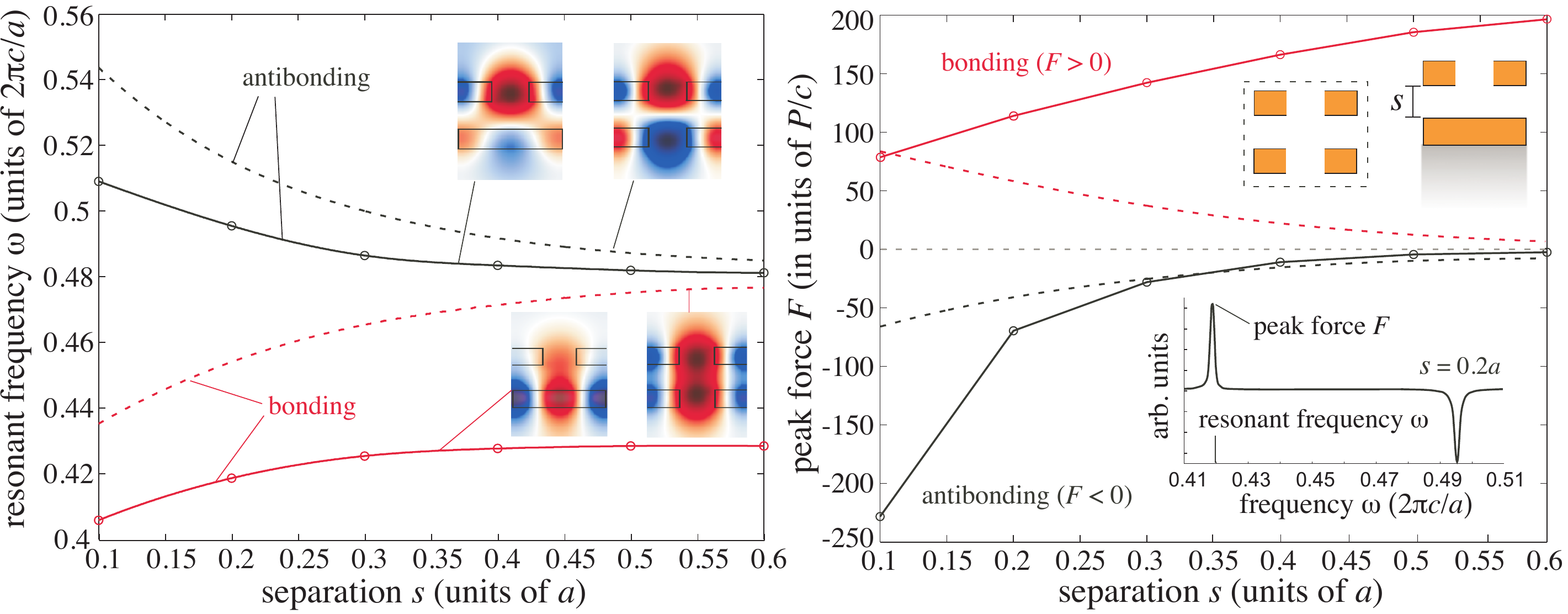}
\caption{(\emph{Left}:) Resonance frequency $\omega$ (units of $2\pi
  c/a$) a function of separation $s$ (units of membrane period $a$),
  for both the single-membrane (asymmetric) structure of \figref{fig1}
  (solid lines) as well as the double-membrane (symmetric) structure
  of~\citeasnoun{Liu09} (dashed lines). The insets show the electric
  field component $E_x$ in the $x$--$z$ plane ($y=0$) near $\omega$ at
  a particular $s=0.3a$. In the symmetric case, the attractive and
  repulsive modes are in-phase and out of phase, respectively, as
  expected. (\emph{Right}:) Resonant (peak) force $Fc/P$ (units of
  incident power $P/c$), at the resonant frequencies $\omega$ plotted
  on the left figure, as a function of $s$.  The bottom inset shows
  the broad-bandwidth force spectrum of the asymmetric structure at a
  particular $s=0.2a$, showing both the bonding ($F > 0$) and
  antibonding ($F < 0$) resonances. The inset also denotes what is
  meant by resonance frequency $\omega$ and peak force $F$.}
\label{fig:fig2}
\end{figure}

In this section, we explore attractive and repulsive resonances in the
asymmetric membrane structure of \figref{fig1}, along with possible
applications and the underlying theory.

\subsection{Symmetric and asymmetric systems}
\label{sec:bonding}

We begin by considering two related membrane structures: the
asymmetric structure of \figref{fig1}, with a perforated silicon
membrane over a silica substrate; and also a symmetrized version
consisting of two identical perforated silicon membranes, examined
previously by~\citeasnoun{Liu09}.  In both cases, the membranes are
illuminated from above by normal-incident ($z$ direction) light
polarized in the $x$ direction, and we consider the resonant
frequencies and the resulting forces as a function of frequency and
separation.

In the symmetric case, it is well known that each resonance of the
individual membranes splits into two resonances of the coupled
two-membrane system: ``bonding'' and ``anti-bonding'' modes, in which
the individual resonances are excited in phase and out of phase,
respectively~\cite{Povinelli05}.  The frequencies of these two
resonances as a function of membrane separation are shown as dashed
lines in the left plot of \figref{fig2}, and as expected the frequency
splitting vanishes as the separation increases (and hence the membrane
coupling decreases)~\cite{Povinelli05}.  The corresponding $E_x$ field
patterns are shown as the right two insets of the left plot, and
display the expected phases.  Each resonant mode corresponds to a
resonant peak in the optical force, and this peak force (at the
resonant frequency) is plotted as a function of separation as dashed
lines in the right part of \figref{fig2}.  As expected, the bonding
and anti-bonding modes correspond to opposite-sign attractive and
repulsive forces between the membranes, respectively, and the force
becomes stronger as the separation decreases (increasing the membrane
interactions).  The attractive force in the bonding case has slightly
larger magnitude than the anti-bonding repulsion, which can be
explained by the larger field overlap in the former case due to the
lack of a node in $E_x$ between the
membranes~\cite{PovinelliJo05,NgChan05}.

In the asymmetric case, the mode of the isolated membrane is not the
same frequency as the corresponding guided mode of the isolated
layered-substrate structure (although the parameters can be adjusted
to force a degeneracy if desired).  The mode of the silicon on silica
(oxide) system is actually a lossless waveguide mode (lifetime $\sim Q
= \infty$), not a resonance; it is only when the membrane is brought
into proximity with the oxide that the membrane's periodicity $a$
allows guided modes at wavevector $2\pi/a$ (and multiples thereof) to
couple to normal-incident radiation~\cite{JoannopoulosJo08-book}.  In
this case, the layered-substrate (waveguide) mode that is nearest in
frequency to the isolated-membrane resonance frequency is the
lowest-order waveguide mode of wavevector $2\pi/a$. Because the two
mode frequencies are no longer degenerate, when the resonant
frequencies of the asymmetric case are plotted versus separation as
solid lines in the left part of \figref{fig2}, the frequency splitting
no longer vanishes as the separation increases.  Nevertheless, there
is a frequency splitting or ``level repulsion,'' explained below in
terms of second-order perturbation theory, which becomes significant
for small separations, and the corresponding field patterns display
the qualitative phase characteristics of bonding/anti-bonding modes
(insets). As a consequence, as considered theoretically below, the
force spectrum in the asymmetric case (right, inset) indeed displays
the characteristic attractive and repulsive resonant peaks of
bonding/anti-bonding modes.  The peak force (at resonance) versus
separation is plotted as solid lines in the right part
of~\figref{fig2}, and has similar sign as in the symmetric case. Of
course, the system is now more complicated than the symmetric case in
a variety of ways (e.g. the field patterns are no longer
symmetrical/anti-symmetrical and the lifetimes as well as the
frequencies depend strongly on separation), so the peak force versus
separation dependence is significantly different: First, the peak
bonding (attractive) force \emph{decreases} as $s$ decreases and
reaches a \emph{constant value} as $s \to \infty$. Second, the ratio
of the antibonding (repulsive) to bonding force becomes increasingly
larger at smaller separations (e.g. it is more than a factor of 2
larger at $s=0.1a$), in contrast to what is normally
observed~\cite{PovinelliJo05,NgChan05}.

To understand these features of the force, we use the fact (reviewed
in \secref{pert}) that the force is proportional to both the
separation ($s$) dependence $d\omega/ds$ of the frequency and also the
lifetime $Q$.  As discussed below, perturbation theory indicates that
$d\omega/ds$ of a nondegenerate mode in one object decreases
proportional to the square of its field overlap with the other
object~\cite{Johnson02:bound}.  For a mode which as $s\to\infty$
approaches a leaky mode of the perforated membrane---in our case, the
antibonding mode---the lifetime $Q$ asymptotes to a nonzero constant
and hence the product $Q d\omega/ds\to 0$; correspondingly, the force
tends exponentially to zero with $s$.  Similarly, the force must tend
to zero for both the bonding and antibonding modes of a symmetric
membrane (where all modes are leaky as $s\to\infty$).  On the other
hand, for a mode that asymptotes as $s\to\infty$ to a lossless guided
mode of the unpatterned substrate---in this case, the bonding
mode---the lifetime $Q$ diverges as $s\to\infty$.  In fact,
perturbative scattering theory~\cite{Johnson02:bound} indicates that
the scattered power, and hence $1/Q$~\cite{JoannopoulosJo08-book}, is
proportional to the square of the field overlap with the periodic
membrane (the source of the scattering loss). Hence $Q$ diverges at
the \emph{same rate} at which $d\omega/ds$ vanishes, and thus the
force should asymptote to a nonzero constant as $s\to\infty$.  These
behaviors are precisely what is observed in~\figref{fig2} (right): the
peak forces of the symmetric system and the asymmetric antibonding
mode decrease monotonically to zero with increasing $s$, while the
peak force of the asymmetric bonding mode increases monotonically to a
constant with increasing $s$. The corresponding variation in $Q$ is
shown in \figref{fig3} (left).  Note that for $s \lesssim 0.2a$, the
$Q$ of the antibonding mode increases rapidly with decreasing $s$,
leading to an increasing antibonding force that is many times larger
than the corresponding bonding force.  In a practical system, the
behavior of the bonding mode will be further modified by the presence
of loss in the isolated-substrate guided mode (from finite-size
effects, roughness, absorption, etcetera)---this will cause its $Q$ to
saturate to a finite value. In this case, the force will behave
\emph{nonmonotonically}: it will initially increase, but will then
decrease to zero as $s$ goes beyond the saturation point of $Q$ (while
$d\omega/ds$ continues to decrease). Thus, the lifetimes of both the
membrane and substrate could be exploited to tailor the $s$ dependence
of the force in this and other similar systems.

\subsection{Tunable mechanical properties}
\label{sec:apps}

\begin{figure}[t!]
\centering
\includegraphics[width=1.0\columnwidth]{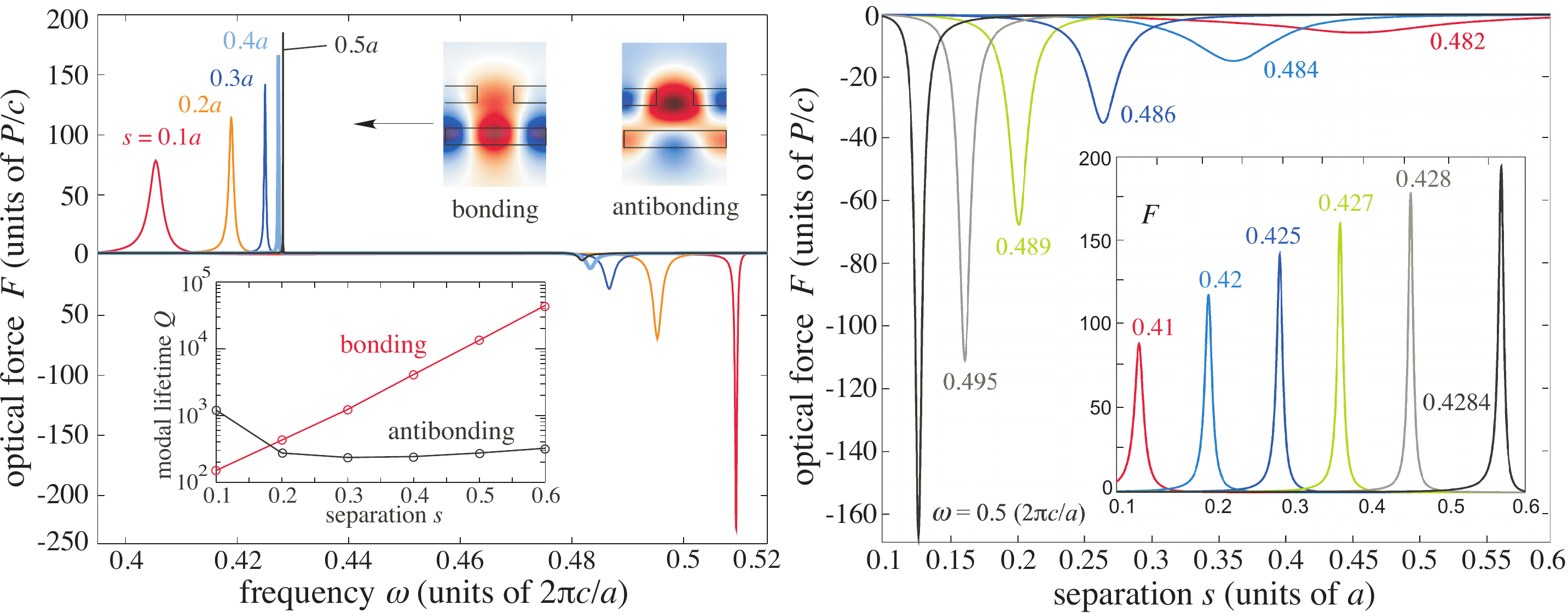}
\caption{(\emph{Left}:) Optical force $Fc/P$ on the single-membrane
  structure of \figref{fig1}, as a function of the frequency $\omega$
  of incident light of power $P$, for various separations $s$. The
  insets show typical $E_x$ field patterns (in the $x$-$z$ plane, at
  $y=0$) for both the attractive (left) and repulsive (right)
  resonances. (\emph{Right}:) Optical force $Fc/P$ as a function of
  separation $s$ for incident light input at various frequencies
  $\omega \in [0.48, 0.5]$~$(2\pi c/a)$.  The bottom inset shows
  $Fc/P$ for light input over a lower frequency range $\omega \in
  [0.41,0.43]$~$2\pi c/a$. The force versus $s$ plot was obtained by
  fitting the force spectrum obtained via FDTD at a few $s$ to a sum
  of Lorentzian resonances, and then interpolating the resulting
  Lorentzian parameters over a denser $s$ range.}
\label{fig:fig3}
\end{figure}

Because the frequency and magnitude of the resonant forces change as
$s$ is varied, it is interesting to study also the separation
dependence of the force for light incident at a single frequency,
which alters the mechanical dynamics of the membrane. Here, we
consider the effect of light incident at a single frequency, and then
extend our analysis to the case of two frequencies. (More generally,
it may prove interesting to study the dynamics of the membrane for
modulated pulses.) \Figref{fig3} plots the optical force $Fc/P$ as a
function of separation $s$ for incident light at various frequencies
$\omega$. As expected, the lifetime~$\sim Q$ of the attractive peak,
with $E_x$ concentrated in the layered substrate (shown on the inset),
becomes infinite ($Q \to \infty$) as $s \to \infty$ due to the reduced
coupling between the infinite-$Q$ substrate mode and the finite-$Q$
PhC resonance. At a fixed frequency, changing $s$ can move the system
into or out of resonance, leading to a dramatic $s$-dependence of the
force.  (Indeed, the $s$ dependence can be much sharper than shown
here, e.g. if the hole diameter is shrunk to increase the $Q$ of the
resonances.) One can obtain transitions in the sign of the force, from
attractive to repulsive and vice versa, as $s$ is varied, leading to
stable and unstable equilibria, not yet including the mechanical
restoring force from the membrane supports, and even multiple
equilibria.

When mechanical forces are included, two things can happen. If the
optical force is nonzero, the mechanical equilibrium point of the
membrane will shift and the slope of the force curve (the spring
constant $\kappa_o = dF/ds$) will be altered.  If one operates at a
point where the optical force is zero, then the equilibrium position
is unaltered but the spring constant is changed. The total spring
constant, including linear mechanical restoring forces on the
membrane, will be given by $\kappa = \kappa_o + \kappa_m$, where
$\kappa_m$ denotes the mechanical spring constant. Whereas $\kappa_m$
is frequency- and power-independent, $\kappa_o$ exhibits a very
sensitive dependence on both, and therefore by choosing $\omega$ and
the incident power it is possible to tune the total spring constant of
the system~\cite{Sheard04,Povinelli04:slow,Mizrahi07,Alegre10}. On the
one hand, if one chooses $\omega$ so that $\kappa_o / \kappa_m > 0$,
then optical forces act to increase $\kappa$ and therefore stiffen the
stable equilibrium. In systems driven by undesirable thermal
fluctuations, this effect has been exploited for ``cooling'' the
resulting vibrations~\cite{Camacho09,Rosenberg09,Thourhout10}. On the
other hand, if one chooses $\omega$ so that $\kappa_o / \kappa_m < 0$,
then $\kappa$ can be decreased and even \emph{flip sign} as the
optical power increases, leading to an unstable equilibrium and
bistable behavior~\cite{Sheard04,Mizrahi07}. Near an exact
cancellation $\kappa_o \approx -\kappa_m$, the linear term in the
$s$-dependence of the force is decreased relative to the higher-order
nonlinear terms (which include both optical and mechanical terms),
allowing arbitrarily strong nonlinear mechanical effects, and even a
strictly nonlinear regime of operation ($\kappa_o = -\kappa_m$) where
effects like bistability, hysteresis, and frequency conversion should
be readily observable~\cite{KippenbergRo05,Kippenberg08}.

\begin{figure}[t!]
\centering
\includegraphics[width=1.0\columnwidth]{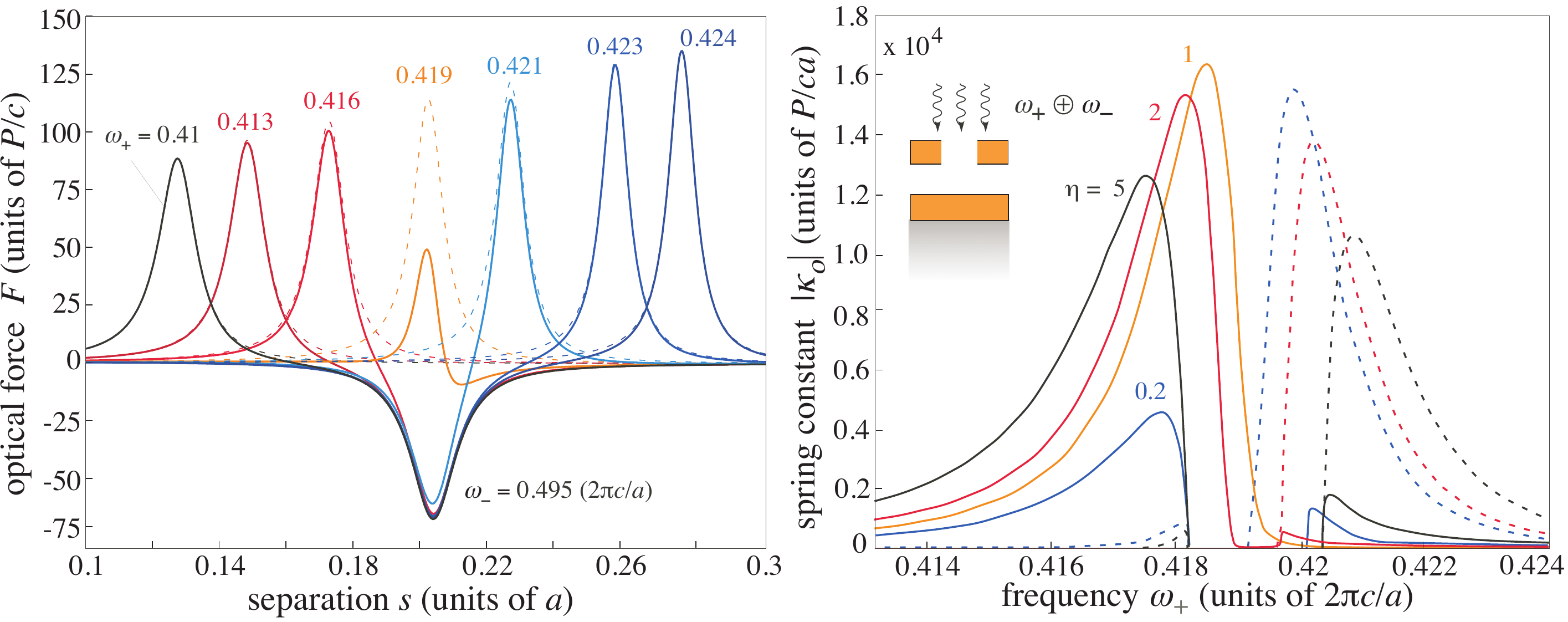}
\caption{(\emph{Left}:) Optical force $Fc/P$ as a function of
  separation $s$, for light incident at two frequencies $\omega_{+}$
  (varied) and $\omega_{-} = 0.495$ $(2\pi c/a)$, with corresponding
  power $P_{+}$ and $P_{-}$, respectively.  Dashed lines show the
  force for $P_{-}=0$. (\emph{Right}:) Absolute value of optical
  spring constant $|\kappa_o|$ (units of $P/ca$) as a function of
  frequency $\omega_+$. Dashed and solid lines correspond to negative
  (unstable) and positive (stable) values of $\kappa_o$, plotted for
  different values of $\eta = P_{-} / P_{+}$.  Both $Fc/P$ and
  $\kappa_o$ are normalized against the total input power $P = P_{+} +
  P_{-}$.}
\label{fig:fig3a}
\end{figure}

For light incident at a single frequency $\omega$, sign transitions in
the force occur as the structure moves toward or past a force
resonance, as illustrated in \figref{fig3}. Away from these
resonances, the ``background'' dimensionless force $Fc/P$ is
attractive and bounded from above by $2$ (see \secref{fp}), leading to
negligible optical spring constants (small $\kappa_o$) at the
corresponding equilibria separations [see \figref{fig3} (inset)]. This
generally does not preclude a strong modification of the mechanical
properties of the membrane (achieving large $\kappa_o \sim \kappa_m$)
since it is also possible to operate at separations where $Fc/P$ is
large and has linear slope ($dF/ds \sim s$), although this inevitably
causes a change in the initial mechanical equilibrium separation of
the membrane~\cite{Kippenberg:07}.  For applications in which
achieving a large $\kappa_o$ without modifying the initial mechanical
separation (i.e. achieving a large $dF/ds$ at a position where $Fc/P =
0$) is important, then a different scheme is required.  For example,
rather than operating with incident light at a single frequency
$\omega$, one can instead consider the combined effect of light
incident at two different frequencies $\omega_+$ and $\omega_{-}$,
near the attractive (bonding) and repulsive (antibonding) resonances,
respectively.  This idea is illustrated in \figref{fig3a} (left),
which shows the optical force $Fc/P$ as a function of separation $s$
for incident light of power $P=P_++P_{-}$ consisting of two
frequencies, $\omega_+$ [chosen in the region $\omega_+ \in
  [0.41,0.424]$~$(2\pi c/a)$] and $\omega_{-} = 0.495$~$(2\pi c/a)$,
of corresponding power $P_{+}$ and $P_{-}$, respectively. From
\figref{fig3}, it is clear that incident light at $\omega_{-}$ leads
to a repulsive peak at $s \approx 0.2a$ whereas incident light in the
$\omega_+$ range leads to attractive resonances in the range $s \in
[0.1, 0.3]a$. As the attractive and repulsive peaks excited by
$\omega_{-}$ and $\omega_+$ come close to one another, the transitions
in the sign of the force become more pronounced, leading to larger
$\kappa_o$. To quantify the enhancement in the spring constant,
\figref{fig3a}~(right) plots the absolute value of the optical spring
constant $|\kappa_o|$ (units of $P/ca$) as a function of $\omega_+$,
for different values of the ratio $\eta = P_{-} / P_{+}$ of power in
$\omega_{-}$ versus $\omega_+$, where dashed/solid lines correspond to
negative (unstable) and positive (stable) $\kappa_o$, demonstrating
orders-of-magnitude enhancement in $\kappa_o$. For example, the peak
spring constant $|\kappa_o|$ in the case where $\eta= 1$ is $\approx
10^4$~$(P/ca)$, whereas it is smaller than $10$~$(P/ca)$ in the case
of incident light only at $\omega_{-}$, corresponding to $\eta \to
\infty$. An alternative scheme that allows tailoring of the optical
spring constant near equilibrium was explored
in~\citeasnoun{Rakich07}, although in that case the effect is achieved
by the presence of multiple bonding/antibonding pairs in which
opposite-sign force resonances were designed to occur at closely
spaced frequencies, whereas here there is no need for the resonances
to be closely spaced.

Additional forces on the membrane arise at small separations due to
residual static charges and also due to quantum/thermal fluctuations
(Casimir and van der Waals forces), which are typically
attractive~\cite{milton04,Capasso07:review,Genet08,Klimchitskaya09}
and may lead to ``stiction'' problems in micromechanical (MEMS)
systems where moving parts are forced into
contact~\cite{Serry98,hochan1,DelRio05}.  Here, the separation
dependence of the classical optical force can potentially be used to
combat such stiction effects~\cite{Ekinci05,Pernice10}.  Not only can
one exploit a repulsive resonance to oppose stiction, but the
separation dependence means that such a repulsion can be designed to
take effect only if $s$ inadvertently falls below some threshold.
That is, a repulsive resonance for small $s$ can be used as a feedback
effect to reduce the chance of stiction without significantly altering
the mechanical dynamics at larger $s$ where the incident light is out
of resonance. In an upcoming manuscript, we will demonstrate how these
effects can also be exploited to design integrated,
all-optical~\footnote{The interaction of normal-incident light with
  the membrane in this system can be exploited to simultaneously
  control and measure the membrane's equilibrium separation.} and
accurate techniques for measuring the Casimir effect that rely on
measuring static displacements rather than forces or force
gradients.

\Figreftwo{fig3}{fig3a} only show a small sample of the kinds of
optical effects that can be observed in evanescently-coupled
systems. In particular, there are many degrees of freedom and
possibilities to explore, especially if one is not restricted to
symmetric structures or operating at a single frequency. For example,
the magnitude of the optical forces shown in \figref{fig3} are by no
means the largest possible, since larger forces can be obtained merely
by increasing $Q$ at the expense of bandwidth (and $s$ insensitivity).
Even more complicated behaviors can be obtained by increasing the
number of resonances, and the choice of resonance offers a
corresponding choice of lengthscales or operating frequencies.

\subsection{Level repulsion in asymmetric membranes}
\label{sec:pert}

For well-defined (long lifetime) resonant modes, perturbation theory
has been used to analyze the relationship between the force and the
resonant frequency/lifetime~\cite{PovinelliJo05,Chan09,Alegre10}, and
this relationship can also be used to illuminate the relationship
between the sign of the force and the field distribution, as well as
the physical origin of resonant repulsion.

As long as the interactions in the system are dominated by a discrete
set of well defined leaky resonances (lifetime $\gg$ 1/period), then
one can apply standard methods of discrete-spectrum time-independent
perturbation theory~\cite{Johnson02:bound}.  (For non-resonant systems
with a continuum of non-localized modes participating at every
frequency, perturbative methods are much more
complicated~\cite{Tannoudji77}.)  In this case, we proceed in three
steps: first, we relate the force $F(s)$ to the $s$-dependence
$d\omega/ds$ of the resonant frequency $\omega$ (generalizing a
previous result~\cite{PovinelliJo05}); second, we connect $d\omega/ds$
to the electric-field distribution of the resonance; third, we
incorporate the level-repulsion effects of nearby resonances via
second-order perturbation theory.  There is one important complication
that arises in resonant systems which does not arise in the lossless
guided modes where $d\omega/ds$ effects were previously derived: the
quality factor $Q$ (lifetime $\sim Q/\omega$) of the resonances may in
general also vary with separation.  \citeasnoun{LiPernice09}
considered the effect on the force due to changes in both $\omega$ and
$Q$ and found generally that at \emph{resonance} (including the
structure considered here), $dQ/ds$ effects have little if any effect
on the force (being higher order in $1/Q$, as we previously suggested
without proof~\cite{PovinelliJo05}). We therefore neglect $dQ/ds$ in
the following, in which case the expression for the force is:
\begin{align}
  \frac{Fc}{P} =
  -\frac{c}{P}\frac{d}{ds}\left(P\frac{Q}{\omega}\right) =
  \frac{Qc}{\omega^2} \frac{d\omega}{ds},
\label{eq:F-omega}
\end{align}
where $F < 0$ denotes a repulsive force.  Although \eqref{F-omega} is
derived rigorously from coupled-mode theory
in~\citeasnoun{LiPernice09}, a simple justification for the same
result can be obtained if one neglects radiative or absorptive losses,
to treat the resonator as a closed system (i.e. equivalent to changing
the separation slowly compared to $1/\omega$ but quickly compared to
the lifetime).  Given an incident power $P$, the energy $U$ stored in
a resonance of real frequency $\omega$ and quality factor $Q$ is (by
definition of $Q$~\cite{JoannopoulosJo08-book}) given by $U =
PQ/\omega$.  Neglecting radiation loss, any change $dU/ds$ in the
energy must be due to a mechanical force $F = -dU/ds$, resulting in
\eqref{F-omega}.  Equivalently, $F/U =1/\omega \,(d\omega/ds)$, a
result we previously derived for guided ($Q=\infty$)
modes~\cite{Povinelli05}.


The dependence of $\omega$ on $s$ can be predicted by perturbation
theory. In particular, the first-order change $\delta \omega^{(1)}$ to
the frequency $\omega$ coming from a small change $\Delta \varepsilon$ in
the permittivity of a system with original permittivity
$\varepsilon$ is readily expressed as:
\begin{align}
  \frac{\delta \omega^{(1)}}{\omega} &= -\frac{1}{2}
  \frac{\bra{\vec{E}_\omega} \Delta \varepsilon
    \ket{\vec{E}_\omega}}{\bra{\vec{E}_\omega} \varepsilon
    \ket{\vec{E}_\omega}}
\label{eq:dw1}
\end{align}
In this case, however, $\Delta\varepsilon$ is not small: at a given
point near the interface, $\varepsilon$ is changing discontinuously as
that interface moves past the point.  In this case, perturbation
theory must be derived more carefully~\cite{Johnson02:bound}.  For an
interface from $\varepsilon_1$ to $\varepsilon_2$ that is moving by
$\Delta s$ (towards $\varepsilon_2$), assuming isotropic materials,
the numerator of \eqref{dw1} changes to:
\begin{align}
  \bra{\vec{E}} \Delta \varepsilon \ket{\vec{E}} &\to
  \bra{\vec{E}_\parallel} \Delta s (\varepsilon_1 - \varepsilon_2)
  \ket{\vec{E}_\parallel} \nonumber \\ - & \bra{\vec{D}_\perp} \Delta
  s \left(\varepsilon^{-1}_1 - \varepsilon^{-1}_2\right)
  \ket{\vec{D}_\perp}
\label{eq:dw1i}
\end{align}
Without loss of generality, we can hold the upper membrane fixed and
move the substrate (or lower membrane) away by $\Delta s$.  From
\eqref{dw1i}, the way to obtain attractive ($d\omega/ds > 0$) and
repulsive ($d\omega/ds < 0$) resonant effects is clear. If we hold one
membrane fixed and move the substrate (or the other membrane),
$d\omega/ds$ will be positive (attractive) if $|\vec{E}|^2$ is peaked
where $\varepsilon_1 < \varepsilon_2$, i.e. on the air/silicon
interface (adjacent to the upper membrane).  Conversely, $d\omega/ds$
will be negative (repulsive) if $|\vec{E}|^2$ is peaked where
$\varepsilon_1 > \varepsilon_2$, i.e. on the silicon/oxide interface
(away from the upper membrane).  Precisely such field patterns can be
observed in the insets of \figref{fig2} and \figref{fig3}: the
repulsive and attractive modes have $E_x$ peaked at the expected
interfaces.  Note that a homogeneous substrate, e.g. semi-infinite
silicon or oxide, has no interface except for the air interface
adjacent to the upper membrane, so in this case a repulsive force
cannot arise by this mechanism, as demonstrated for the $h_2=0$ case
on the inset of \figref{fig4} below.  An exception to this rule is
discussed in \secref{fp}, in which repulsive forces arising from
radiative modes are analyzed, a situation where a lack of
normalizability causes the perturbation theory to break down.

The above discussion indicates which field patterns might be expected
to lead to repulsion and attraction, but does not explain how such
field patterns can arise.  For the case of a symmetric membrane,
symmetry considerations predict that the degenerate modes of two
isolated membranes will split into even/odd bonding/anti-bonding pairs
by degenerate first-order perturbation
theory~\cite{Messiah76:pert}. The presence of a nodal plane bisecting
the anti-bonding mode (assuming its field is dominated by $E_{xy}$ and
not $E_z$) means that the field pattern will be stronger on the far
sides of the membranes, leading to a repulsive interaction as observed
in \figref{fig2} and \figref{fig3}, and as predicted above.  For
asymmetric membranes, however, there is typically no degeneracy, and
the corresponding two-mode interaction must instead be analyzed by
second-order perturbation theory, which plays a role for sufficiently
small separations (large interactions).  When two isolated waveguides
each have a mode with nearby frequencies, and one brings the
waveguides together so that the mode fields overlap, second-order
perturbation theory predicts a contribution to $\Delta \omega$ that
tends to split the two frequencies:
\begin{equation}
  \frac{\delta \omega^{(2)}}{\omega} = \frac{1}{4}
  \frac{|\bra{\vec{E}_\omega}\Delta \varepsilon
    \ket{\vec{E}_\omega}|^2}{|\bra{\vec{E}_\omega}\varepsilon
    \ket{\vec{E}_\omega}|^2} \nonumber - \frac{1}{2}\sum_{\omega'
    \neq \omega} \left(\frac{\omega^3}{\omega'^2 - \omega^2}\right)
  \frac{|\bra{\vec{E}_{\omega'}}\Delta \varepsilon
    \ket{\vec{E}_\omega}|^2}{\bra{\vec{E}_{\omega'}} \varepsilon
    \ket{\vec{E}_{\omega'}} \bra{\vec{E}_\omega} \varepsilon
    \ket{\vec{E}_\omega}}
\label{eq:dw2}
\end{equation}
(As above, the overlap integrals are modified into $\vec{E}_\parallel$
and $\vec{D}_\perp$ components for motion of discontinuous
interfaces~\cite{Johnson02:bound}.)  Note that the $\omega'^2
-\omega^2$ term pushes $\omega$ away from $\omega'$, and becomes
stronger as the frequencies become closer (with the most dramatic case
being degenerate modes, where the derivation is modified).  Because of
the competition between the first- and second-order terms, which may
be comparable in magnitude for small separations (large overlaps), it
is possible for the force near a resonance to switch signs with
separation, a possibility that is demonstrated in~\secref{multi-mode}
below.

\subsection{Multi-modal interactions}
\label{sec:multi-mode}

Previously, we considered resonances at a relatively low frequency
(compared to $2\pi c/a$), where the only relevant interactions were
between two modes (one for each isolated membrane or substrate).  At
higher frequencies, the density of states generally increases, and
thus many more resonant modes are typically present.  Correspondingly,
the inter-modal interactions become more complicated, and it is not
always possible to identify individual pairs of bonding/anti-bonding
modes.  However, the qualitative features of repulsive and attractive
resonances are still present, although the additional complexity
provides more degrees of freedom leading to more complicated force
phenomena.

\begin{figure}[t!]
\includegraphics[width=0.99\columnwidth]{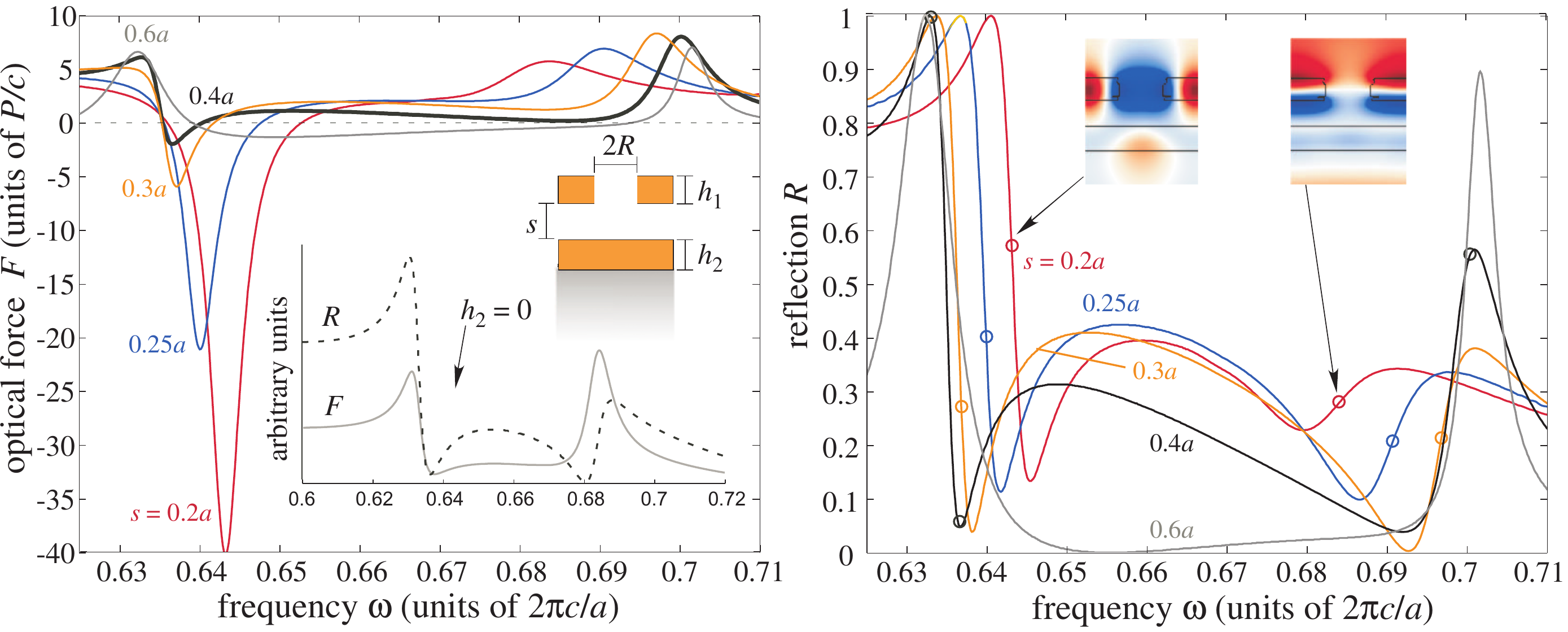}
\caption{(\emph{Left}:) Optical force $Fc/P$ as a function of
  frequency $\omega$ for light of power $P$ incident on the
  single-membrane structure of \figref{fig1}, for various separations
  $s$.  The bottom inset shows the force (solid lines) and reflection
  (dashed line) of the same geometry but for $h_2 = 0$ and
  $s=0.3a$. (\emph{Right}:) Corresponding reflection spectrum $R$ as a
  function of $\omega$.  The open circles indicate frequencies for
  which there exist force minima or maxima. The insets show the
  electric field component $E_x$ in the $x$--$z$ plane ($y=0$) at a
  particular $s=0.2a$, and at the indicated frequency points $\omega =
  0.64 (2\pi c/a)$ (left) and $\omega = 0.681 (2\pi c/a)$ (right).}
\label{fig:fig4}
\end{figure}

For example, the force and reflection spectra for the asymmetric
membrane system of \figref{fig1} are shown in \figref{fig4} in a
higher frequency window (about double the frequencies in
\figref{fig3}).  As before, the force spectrum (left) shows both
repulsive and attractive resonances.  In this case, however, we
actually observe a force resonance changing \emph{sign} as a function
of separation, which physically can be interpreted as different terms
dominating in the perturbation theory [\eqref{dw2}] at different
separations.  In the reflection spectrum (right), these resonances
correspond to Fano shapes (adjacent peaks and dips), a well-known
consequence of the coherent combination of a resonant process with
direct transmission through the slabs~\cite{Fan03:CMT,suh05,Liu09}.
Because the reflection spectrum depends sensitively on the separation,
the peak locations from a broad-bandwidth low-intensity source could
be used to accurately determine the separation in an experiment.  Note
also that the resonant peaks in this frequency window approach one
another as the separation \emph{decreases}, which means that the
largest contribution to level repulsion in this case is coming from
interactions with other modes (outside this frequency window, not
shown); this is verified by examining the field patterns (upper
insets), which clearly correspond to completely different modes in the
membrane and not just a relative-phase change. As in \secref{apps},
fixing the frequency and plotting the force versus separation reveals
a force that changes both magnitude and sign as a function of
separation.



In the absence of a guided mode in the substrate, the physics of this
situation is completely changed---one no longer has level repulsion
effects, since there are only resonances in the membrane. A general
argument for the attractive nature of the resonant force is given in
the previous section. This occurs, for example, if the oxide substrate
is replaced simply by a low-index (oxide, $n=1.5$) substrate, which
supports no guided modes of any kind on its own.  However, the
resonant frequencies (and lifetimes) of the membrane are still
modified by the proximity of the substrate, so there is still a
resonant force, and in this particular geometry we find that the
resonant force is always attractive. Such attractive forces can
however be exploited in integrated photonic devices as a way of tuning
the mechanical response of the
devices~\cite{LiPernice08,MaPovinelli10}. The force and reflection
spectra for this case are plotted in the middle inset of \figref{fig4}
(left), for a particular separation $s=0.3a$.

\section{Fabry--Perot Forces}
\label{sec:fp}

Our focus thus far has been on periodic structures supporting
resonances whose coupling leads to both enhanced attractive \emph{and}
repulsive forces.  However, at large separations compared to the
evanescent tail of the participating guided modes, evanescent
bonding/anti-bonding effects lead to negligible force enhancement. On
the other hand, there exists an alternative and complementary
force-enhancement mechanism that can play a role at large separations:
light normally incident on two separated planar objects can be
resonantly enhanced due to the Fabry--Perot cavity formed
\emph{between} the objects by reflections from adjacent surfaces of
the objects, and the extent of this enhancement will increase with the
reflectivity of the objects (i.e. with the $Q$ of the
cavity). Although such Fabry--Perot force-enhancement mechanisms have
been considered in previous
work~\cite{Braginskii77,Dorsel83,KippenbergRo05,Mizrahi06,Regal08,Rosenberg09},
in applications ranging from gravitational-wave
detection~\cite{AlexAbramovici92} to optical
cooling~\cite{Schliesser06,Gigan06,Kleckner06,Arcizet06,LinRo09}, it
is interesting to explicitly compare the two resonant mechanisms.  In
this section, we briefly consider the kinds of repulsive and
attractive forces that can arise in systems consisting of
\emph{unpatterned} multilayer objects, emphasizing some of their
similarities and differences compared to forces arising from
evanescently coupled resonances.

\begin{figure}[t!]
\centering
\includegraphics[width=0.5\columnwidth]{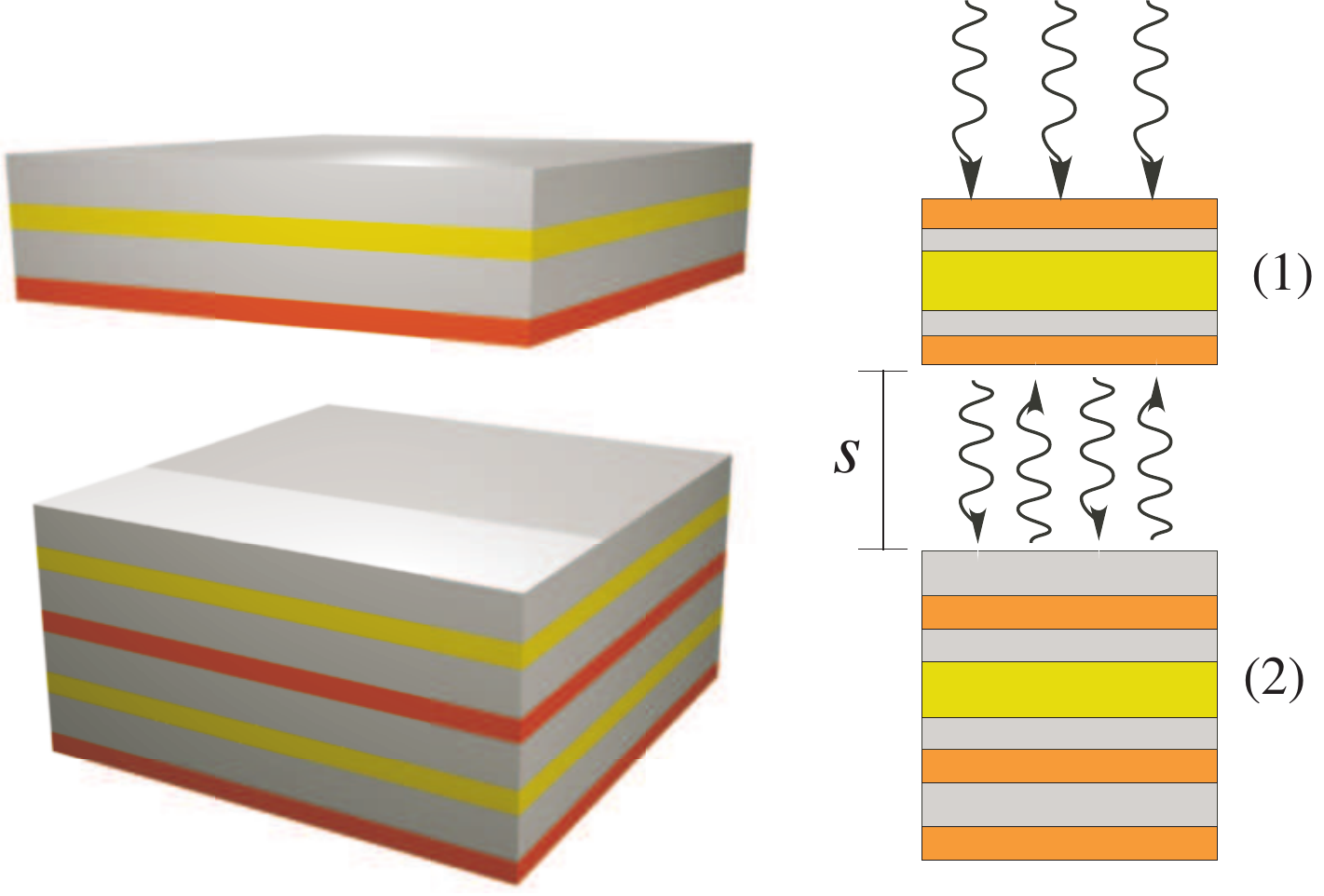}
\caption{Schematic of system consisting of two multilayer objects
  [labeled as (1) and (2)] separated by a distance $s$. A
  two-dimensional cross-section for the particular case of two
  quarter-wave stack mirrors with a defect (yellow) is shown on the
  right.}
\label{fig:fig6}
\end{figure}

Resonant radiation pressure within Fabry--Perot cavities has a long
history, dating back to work in the 1960s on interferometer
sensitivity~\cite{Braginsky67}, and has since been considered both
theoretically and experimentally for many
applications~\cite{Mizrahi06,Kippenberg08}, such as nonreciprocal
phenomena~\cite{Manipatruni09}, optical
cooling~\cite{Gigan06,Arcizet06,Kleckner06}, and tunable optical
springs~\cite{Corbitt07}.  If the space between two partially
reflecting mirrors (e.g. Bragg mirrors) is viewed as a waveguide, then
the resonance frequency for normal-incident light corresponds to a
slow-light (zero group-velocity) band edge where radiation pressure is
enhanced~\cite{Povinelli04:slow,Mizrahi06}.  In all of these cases,
the pressure is repulsive, as one might expect for light bouncing
between the two objects (and was argued in general for two
semi-infinite objects~\cite{AntonoyiannakisPe10}).  We show this in
general below, for radiating modes (not guided modes) and any
unpatterned multi-layered structure (in the absence of gain).

Here, we consider a general class of geometries, depicted in
\figref{fig6}, consisting of two unpatterned (translationally
invariant in two directions) planar multilayer objects separated by
distance $s$ in vacuum, denoted as objects~(1) and~(2), characterized
by complex reflection/transmission amplitudes $r_1/t_1$ and $r_2/t_2$
at a given frequency, respectively (satisfying $|r_k|^2 + |t_k|^2 = 1$
in the absence of absorption or gain). The dimensionless force $Fc/P$
on object~(1) due to light normally incident from above at frequency
$\omega$ can be readily computed (using a simple transfer-matrix
analysis to obtain the stress tensor~\cite{Yeh88}) to be:
\begin{align}
  \frac{Fc}{P} &= 1 + \left| r_1 + t_1 r_2 e^{2i\delta} F_+\right|^2 - |F_+|^2,
\label{eq:Ffb}
\end{align}
where $F_+ = t_1 / [1 - r_1 r_2 \exp(2i\delta)]$ is the induced field
at the lower interface of object (1), and $\delta \equiv 2\pi \omega
s$ is the phase associated with the air gap.

Elementary manipulation of \eqref{Ffb} shows that the force is bounded
above by $Fc/P\leq 2$, which is a bound on the attractive (positive)
force.  This result is a physical consequence of conservation of
momentum: the light trapped between the objects can only act to repel
them, whereas an attractive force can only arise from reflections from
the uppermost surface.  (The key difference compared to patterned
membranes or guided modes is that the fields in between the objects
are now purely propagating waves in which the field amplitude is
proportional to the wave momentum~\cite{Jackson98}, with no evanescent
component where these two quantities can be decoupled.)  The maximum
reflectivity is~100\%, corresponding to $Fc/P=2$.  A related situation
is one in which the lower object plays no role because its
reflectivity approaches zero. In this case, for $r_2=0$ and $|t_2|=1$,
one recovers a standard result for the force on a single object, $Fc/P
= 1 + |r_1|^2 - |t_1|^2$~\cite{Braginsky67}, which is always positive
(attractive) and is bounded above by~2.  (Note, however, that all of
these limits only apply in the absence of gain, which can alter the
force by emitting additional photons from the
objects~\cite{Mizrahi10}.) On the other hand, the repulsive forces are
unbounded, becoming arbitrarily large as $|r_1|$ and $|r_2|$ approach
unity.  This is simply a consequence of the repulsive pressure from
the Fabry--Perot mode trapped between the objects, whose lifetime (and
energy density) diverge in this limit.  This familiar result has been
exploited in many cavity-enhanced optomechanical systems as noted
above.

One can construct multilayer objects supporting exponentially
localized resonances which couple to normally incident radiation.  For
example, this is the case if each object consists of a multilayer
Bragg mirror with an embedded defect
layer~\cite{JoannopoulosJo08-book}.  In this case, degenerate
perturbation theory implies that two symmetric objects [such as those
  in \figref{fig6}~(right)], each with an identical embedded
defect/resonance, should couple to form bonding/anti-bonding states
where the resonances are in/out-of phase.  Naively, one might suppose
that this will lead to repulsive and attractive resonances, as in the
coupled guided-mode case, but \eqref{Ffb} indicates that this is
impossible.  The explanation is straightforward: although such defect
modes are exponentially localized within the Bragg mirrors composing
each object, they are propagating in the region between the objects
where there are no mirrors (because the input beam is propagating in
free space and no diffraction occurs).  This means that the frequency
splitting does not depend exponentially on the separation between the
two objects, and hence there is no resonant force enhancement via
these modes, by the analysis of \secref{forces}.

\section{Conclusion}
\label{sec:conclusion}

Optomechanical interactions are a rich subject of current research,
and the use of evanescently coupled guided resonances enables an
especially rich set of phenomena because of the presence of both
attractive and repulsive resonances. The ability to tailor and exploit
guided resonances coupled via periodic modulations offers an exciting
opportunity to procure complicated force effects at small
separations. In this paper, we showed that functionality similar to
that of previously studied symmetric-membrane systems can be obtained
in asymmetrical membrane-substrate structures.  From an experimental
point of view, such asymmetrical structures are attractive in that
only a single membrane need be suspended and patterned.  From a
theoretical viewpoint, because the resonant modes of asymmetrical
structures need not come in degenerate pairs (unless degeneracies are
forced), more than one pair of modes can have strong interactions,
leading to the possibility of richer force phenomena~\cite{Rakich07}.
Correspondingly, the distance dependence of a force spectrum with
multiple attractive and repulsive resonances can exhibit richer
``optical spring'' phenomena than are possible with repulsive
resonances alone (as in Fabry--Perot resonances between mirrors); for
example, one can operate at a zero of the optical force to tune the
optical spring constant (in either direction) without altering the
equilibrium position.

\section*{Acknowledgements}

We are grateful to Aristeidis Karalis and Peter Bermel at MIT for
useful discussions. This work was supported by the Army Research
Office through the ISN under Contract No. W911NF-07-D-0004, and by the
Defense Advanced Research Projects Agency (DARPA) under contract
N66001-09-1-2070-DOD.

\end{document}